\documentclass[runningheads]{llncs}
\usepackage[T1]{fontenc}
\usepackage{graphicx}
\usepackage{orcidlink}
\usepackage{booktabs}
\usepackage{amssymb}
\usepackage{pifont}
\usepackage{hyperref}
\usepackage{amsmath}
\begin{document}
\title{Privacy Preserving Chest X-ray Classification in Latent Space with Homomorphically Encrypted Neural Inference}
%
%
\author{Jonghun Kim\inst{1,2} \orcidlink{0009-0002-2790-2090} \and
Gyeongdeok Jo \inst{1,2} \and
Sinyoung Ra \inst{3} \and
Hyunjin Park\inst{1,2} \thanks{Corresponding Author} \orcidlink{0000-0001-5681-8918}}
%
\authorrunning{Kim et al.}

%
\institute{Department of Electrical and Computer Engineering, \\ Sungkyunkwan University, Suwon, Korea \and
Center for Neuroscience Imaging Research, \\ Institute for Basic Science, Suwon, Korea \and
Department of Artificial Intelligence, Sungkyunkwan University, Suwon, Korea \\
\email{\{iproj2,oonn1219,nsy0527,hyunjinp\}@skku.edu}}
%
%
%
\newcommand{\xmark}{\ding{55}}

\maketitle              

\begin{abstract}
Medical imaging data contain sensitive patient information requiring strong privacy protection. Many analytical setups require data to be sent to a server for inference purposes. Homomorphic encryption (HE) provides a solution by allowing computations to be performed on encrypted data without revealing the original information. However, HE inference is computationally expensive, particularly for large images (e.g., chest X-rays). In this study, we propose an HE inference framework for medical images that uses VQGAN to compress images into latent representations, thereby significantly reducing the computational burden while preserving image quality. We approximate the activation functions with lower-degree polynomials to balance the accuracy and efficiency in compliance with HE requirements. We observed that a downsampling factor of eight for compression achieved an optimal balance between performance and computational cost. We further adapted the squeeze and excitation module, which is known to improve traditional CNNs, to enhance the HE framework. Our method was tested on two chest X-ray datasets for multi-label classification tasks using vanilla CNN backbones. Although HE inference remains relatively slow and introduces minor performance differences compared with unencrypted inference, our approach shows strong potential for practical use in medical images. Our code is available at \href{https://github.com/jongdory/Latent-HE}{github.com/jongdory/Latent-HE}.

\end{abstract} 
\section{Introduction}

Medical imaging data contain sensitive personal patient information and therefore require strict confidentiality \cite{andriole2014security}. The analysis and processing of medical images often involve sending data to cloud servers or external computing resources, raising important concerns about data security and privacy \cite{mehrtak2021security}. One method to protect privacy when sending data to a server is to encrypt the data before transmission \cite{rivest1978method,aumasson2024serious}. Although this approach eliminates the risk of data leakage, it also requires the encryption and decryption keys to be shared between the server and client. If the secret key of the server is exposed, all the encrypted data can be decrypted \cite{aumasson2024serious}. In addition, because the server must decrypt the data to process them, the server can access the information, posing a privacy risk and possibly leading to a reluctance to send data to the server. 

\begin{figure} [t]
    \centering
    \includegraphics[width=0.95\textwidth]{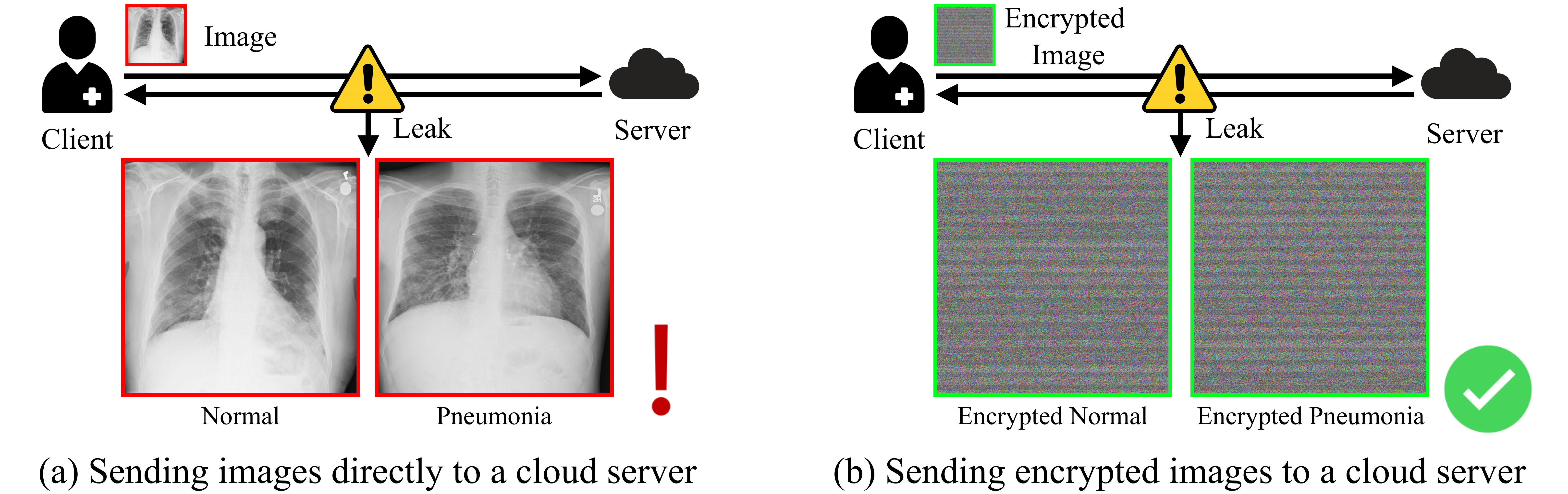}
    \caption{Examples of chest X-ray analysis. (a) Client sends images directly to a server with security risks. (b) Client sends encrypted images to a server without security risks. }
    \label{fig1}
\end{figure}

One solution is to use homomorphic encryption (HE) \cite{gentry2011implementing,brakerski2014leveled,fan2012somewhat,marcolla2022survey}. HE allows data to be processed while still encrypted, thereby allowing operations to be performed without decryption. The results remained encrypted and were identical to those obtained from the original decrypted data. This ensures that the data remain confidential during processing. With HE, the server does not need to know the secret key and the client uses only a public key to encrypt the data. The server can then perform the necessary operations on the encrypted data without accessing the actual information, thereby maintaining data privacy. \textbf{This also means that even if the data is exposed during processing or transmission, the privacy of the information is still protected.} Fig. \ref{fig1} compares the risks of sending unencrypted and encrypted data to a cloud server for medical image analysis.

There are two main HE schemes: Boolean and arithmetic. Torus Fully HE (TFHE) is a well-known Boolean operation scheme \cite{chillotti2020tfhe}. With extensions such as programmable bootstrapping, TFHE can perform some basic arithmetic operations; however, it is primarily designed for bit-level tasks. Complex arithmetic involving integers or real numbers requires a cascade of Boolean circuits, which increases computation time. However, Boolean schemes are efficient for simple logic operations (e.g., AND and OR), particularly for short bit lengths or binary logic. Cheon-Kim-Kim-Song (CKKS) \cite{cheon2017homomorphic} is an HE scheme that supports approximate arithmetic. This allows the direct addition and multiplication of encrypted real numbers. Because CKKS approximates the encryption of floating-point numbers, it allows for a small margin of error. Managing a wide range of precision in CKKS requires larger encryption parameters, which significantly slow the calculation \cite{cheon2017homomorphic}. CKKS is commonly used in statistics and machine learning \cite{lee2022low,hesamifard2017cryptodl,crockett2020low}; therefore, it was used in this study. Inference becomes computationally expensive in the HE framework; thus, there is active research on HE inference \cite{lou2021hemet,lee2022low,ran2023spencnn}. These previous studies only tested HE inference on relatively small images, such as  $32\times32$ images, owing to inference time limitations. In real-world applications, medical images are often much larger, making it difficult to apply these methods directly

To overcome these challenges, this study proposes a framework for medical image inference in HE. First, the images were compressed into smaller latent representations. Subsequently, a classification model was trained on these latent images for analysis. In addition, we adapted efficient computational modules, such as squeeze and excitation blocks \cite{hu2018squeeze}, which have been effective in improving traditional CNNs for the HE framework. To the best of our knowledge, our study is \textbf{one of the first to apply medical image classification of X-ray within the HE framework}, demonstrating that HE can be applied to medical image analysis in which images are large and thus require large parametric models.

\begin{figure} [t]
    \centering
    \includegraphics[width=\textwidth]{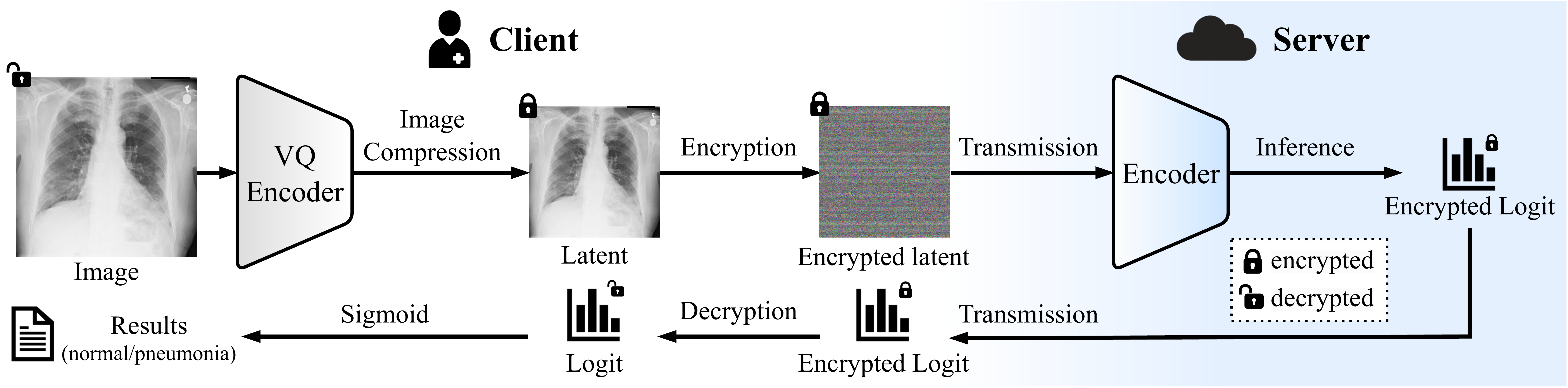}
    \caption{Illustration of the framework for HE inference.}
    \label{fig2}
\end{figure}

\section{Method}
In this paper, we present a framework for HE inference that can be used in the medical image analysis of X-ray images. Fig. \ref{fig2} shows the framework for performing HE inferences on a server using an existing classification model. First, only the public key is shared with the server, while the private key remains with the client. On the client side, the image is compressed into smaller latent representations using image compression and then encrypted with a public key before being sent to the server. The server processes the encrypted data without decryption and generates encrypted logits. These encrypted logits are then sent back to the client, where the client decrypts them using a private key and applies a sigmoid function to obtain the final classification results. The background of HE operations is provided in Section \ref{sec2.1}. Image compression is explained in Section \ref{sec2.2}, and the design of the model architecture is described in Section \ref{sec2.3}.

\subsection{Background}
\label{sec2.1}
HE allows computations to be performed on the encrypted data without decryption. After performing operations on the encrypted data and decrypting, the results are the same as if the operations were performed on the original unencrypted data. Specifically, for the encryption and decryption functions  $\mathbb{E}$ and $\mathbb{D}$ for two data points $x_1$ and $x_2$, 
\begin{equation} 
x_1+x_2 = \mathbb{D}(\mathbb{E}(x_1)+\mathbb{E}(x_2)), \ \  x_1\times x_2 = \mathbb{D}(\mathbb{E}(x_1)\times \mathbb{E}(x_2)). 
\label{eq:1}
\end{equation}
Because computations can be performed directly on encrypted data, sensitive information can be processed on a server while remaining encrypted.

\begin{figure} [t]
    \centering
    \includegraphics[width=\textwidth]{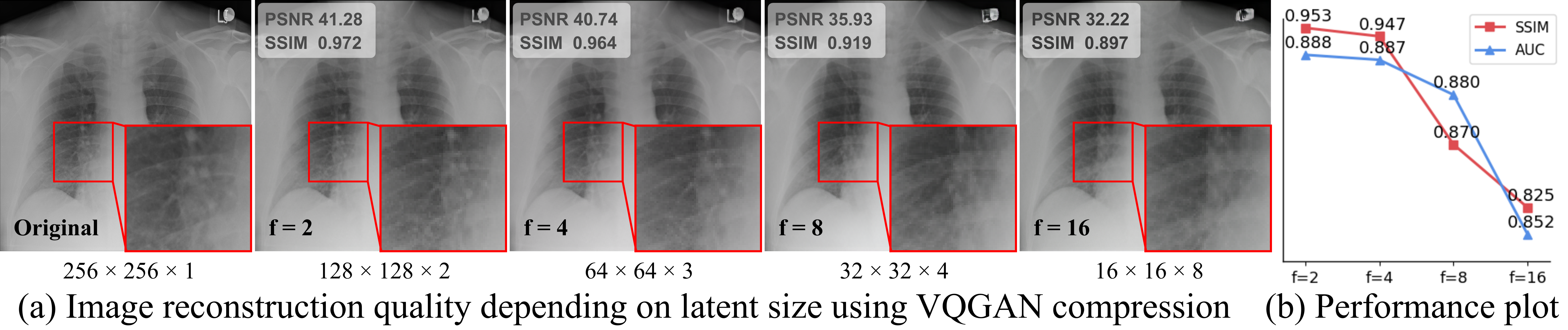}
    \caption{Image quality, model performance, and HE inference time as a function of latent size. (a) Red bounding boxes indicate areas of interest with enlarged views. (b) Performance plot based on the ResNet 20 model according to the downsampling factor.}
    \label{fig3}
\end{figure}

\subsection{Image Compression}
\label{sec2.2}
HE inference is computationally intensive, and inference times increase exponentially with the image size. To address this challenge, this study uses a VQGAN \cite{esser2021taming} to compress images into latent representations. The VQGAN is an effective compression method that generates high-quality latent as shown in recent generative models, such as taming transformers \cite{esser2021taming} and latent diffusion models \cite{rombach2022high}. In our approach, images are first compressed into latent using the VQGAN, and classification tasks are then performed on these latent using a classifier. An example of a reconstructed compressed latent is shown in Fig. \ref{fig3} (a). As expected, there is a trade-off between the size of the latent representation and image quality: higher compression rates lead to lower quality in the reconstructed images. The loss of image quality is also associated with information loss in the latent space, which directly affects classifier performance. As shown in Fig. \ref{fig3} (b), both the image quality and model performance decreased as the downsampling factor $f$ increased. However, reducing the latent size leads to a more efficient inference. This balance among compression, quality, and performance is crucial for optimizing HE-based inferences in medical image analysis. Given an image $x$, the latent representation $z$ is obtained as follows: 
\begin{equation} 
z = \textbf{q}(E(x)), \ \  \hat{x} = G(z),
\label{eq:2}
\end{equation}
where \textbf{q} denotes the quantization function, $E$ is the encoder, $G$ represents the decoder (generator), and $\hat{x}$ refers to the reconstructed image. Our VQGAN was trained using the following loss terms.
\begin{equation} 
\mathcal{L}_{vqgan} = \underbrace{||x - \hat{x}||^2}_{\text{Reconstruction}} + \ \ \underbrace{\text{log}D(x) + \text{log}(1-D(\hat{x}))}_\text{GAN} \ \ +  \underbrace{\mathcal{L}_{vq}}_\text{Codebook}, 
\label{eq:3}
\end{equation}
where $D$ denotes the discriminator and $\mathcal{L}_{vq}$ is the quantization loss \cite{van2017neural}. In this study, we used latent z as the input to the classifier.

\begin{figure} [t]
    \centering
    \includegraphics[width=\textwidth]{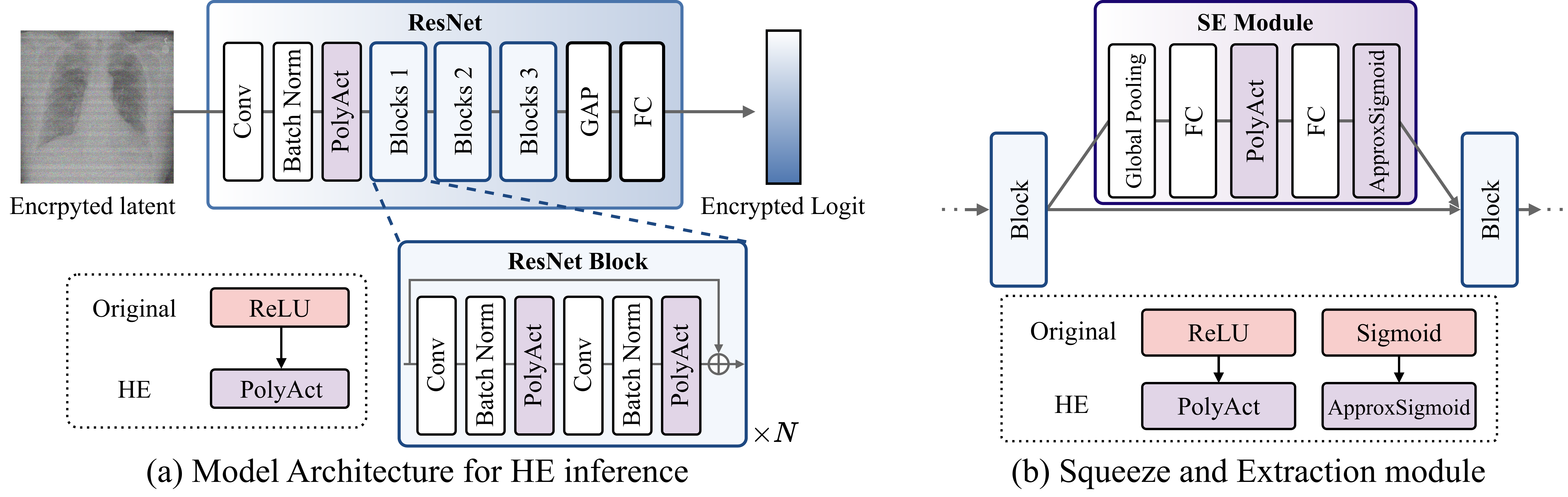}
    \caption{Changes to activation functions from the existing model for HE inference. }
    \label{fig4}
\end{figure}

\subsection{Architecture Design}
\label{sec2.3}\
HE in the CKKS scheme supports only addition and multiplication and not division or logical operations. Therefore, using activation functions such as sigmoid and ReLU is difficult. One solution is to approximate these activation functions using polynomial functions that only require addition and multiplication \cite{peng2022polympcnet,ran2023spencnn}. We adopted this approach in the present study. Fig. \ref{fig4} (a) shows the modified structure of ResNet with the modified activation functions for implementation in HE. The activation function implemented using polynomials for the ReLU, polyact, is defined as follows:
\begin{equation} 
\text{polyact}(x) = a x^2 + b x + c,
\label{eq:4}
\end{equation}
where $a$, $b$, and $c$ are learnable parameters.
Additionally, channel attention modules such as squeeze and excitation (SE) \cite{hu2018squeeze}, have significantly improved the classification performance of previous CNNs. The SE module efficiently improves the performance without significantly increasing the number of parameters. Therefore, implementing this module in HE with limited computational resources can improve the performance with a small increase in the number of parameters Fig. \ref{fig4} (b) shows the modified SE module. The existing SE module uses ReLU and sigmoid functions, which we have approximated with polynomial functions. Specifically, the sigmoid function was approximated as follows:
\begin{equation} 
\text{approxsigmoid}(x)=\alpha x^3 + \beta x^2 + \gamma x + d
 \label{eq:5},
\end{equation}
where $\alpha$, $\beta$, $\gamma$, and $d$ are all learnable parameters. Higher degrees of accuracy can be achieved when approximating activation functions using higher-degree polynomials. However, in HE, each additional multiplication exponentially increases the computational cost, rendering high-degree polynomials impractical \cite{lou2019she}. Therefore, lower-degree polynomials are preferred. However, because a sigmoid module requires a more accurate approximation than ReLU, we use a cubic polynomial for the sigmoid, as in previous studies \cite{hesamifard2017cryptodl,crockett2020low}. 

\section{Experiments}
\noindent \textbf{Implementation Details.} For model training, we used PyTorch v2.0.1 \cite{paszke2019pytorch}, and for HE inference, we used TenSEAL \cite{benaissa2021tenseal} based on Microsoft SEAL \cite{sealcrypto}. The training parameters were set with a batch size of 128 and a learning rate of $2\times10^{-4}$ using the Adam \cite{DBLP:journals/corr/KingmaB14} optimizer. 
We conducted multi-label classification by performing binary classification for each class and compared our models with baseline models including LeNet \cite{lecun1998gradient}, HefNet \cite{ran2023spencnn}, VGGNet \cite{simonyan2015very}, and ResNet (20, 32, 44, 56) \cite{he2016deep}. The model parameters are shown in fig. \ref{fig5} (c). All models were trained and evaluated with a downsampling factor $f=8$, considering the computational cost. The codebook dimension $\mathcal{Z}$ of VQGAN was set to 1024.

\vspace{3pt}
\noindent \textbf{Datasets.} For model training and evaluation, we used CheXpert \cite{irvin2019chexpert} and NIH \cite{wang2017chestxray} chest X-ray datasets, resizing all original images to 256×256 pixels. Both datasets include multilabel classifications for each image. CheXpert, labeled using VisualChexbert \cite{jain2021visualchexbert}, consists of 191,027 subjects, 189,116 for training and 1,911 for testing, with 14 classes. The NIH contains 112,120 subjects, of which 111,010 were used for training and 1,110 were used for testing with 15 classes.

\begin{table} [t]
    \setlength{\tabcolsep}{2pt} 
    \centering
    \caption{Performance results depending on model architecture for chest X-ray classification. \textbf{Micro}, \textbf{Macro}, and \textbf{Weight} represent the micro, macro, and weighted average, respectively. All models in the table were trained and evaluated at $f=8$.}
    \label{table1}
    \centering
    \scalebox{0.7}{
        \begin{tabular}{l|cccccc|cccccc}
            \toprule
            \textbf{Dataset} & \multicolumn{6}{|c}{\textbf{CheXpert}} & \multicolumn{6}{|c}{\textbf{NIH}} \\
            \cmidrule(lr){1-1} \cmidrule(lr){2-7} \cmidrule(lr){8-13}
            \textbf{Metric} & \multicolumn{3}{|c}{\textbf{AUROC} $\uparrow$}  & \multicolumn{3}{c}{\textbf{F1} $\uparrow$} & \multicolumn{3}{|c}{\textbf{AUROC} $\uparrow$} & \multicolumn{3}{c}{\textbf{F1} $\uparrow$} \\
            \cmidrule(lr){1-1} \cmidrule(lr){2-4} \cmidrule(lr){5-7} \cmidrule(lr){8-10} \cmidrule(lr){11-13}
            \textbf{Model} & \textbf{Micro} & \textbf{Macro} & \textbf{Weight.} & \textbf{Micro} & \textbf{Macro} & \textbf{Weight.} & \textbf{Micro} & \textbf{Macro} & \textbf{Weight.} & \textbf{Micro} & \textbf{Macro} & \textbf{Weight.} \\
            \cmidrule(lr){1-13} 
            LeNet         & 0.892 & 0.835 & 0.850 & 0.766 & 0.606 & 0.741 & 0.866 & 0.693 & 0.682 & 0.443 & 0.047 & 0.305 \\
            LeNet + SE  & 0.903 & 0.859 & 0.870 & 0.784 & 0.678 & 0.774 & 0.864 & 0.670 & 0.688 & 0.429 & 0.046 & 0.300 \\
            \cmidrule(lr){1-13}
            VGGNet        & 0.901 & 0.856 & 0.868 & 0.781 & 0.648 & 0.764 & 0.856 & 0.612 & 0.649 & 0.396 & 0.044 & 0.283 \\
            VGGNet + SE   & 0.905 & 0.862 & 0.875 & 0.788 & 0.677 & 0.779 & 0.855 & 0.654 & 0.658 & 0.413 & 0.047 & 0.293 \\
            \cmidrule(lr){1-13}
            HefNet        & 0.909 & 0.869 & 0.879 & 0.796 & 0.695 & 0.787 & 0.856 & 0.658 & 0.674 & 0.425 & 0.046 & 0.299 \\
            HefNet + SE   & 0.911 & 0.870 & 0.883 & 0.798 & 0.694 & 0.790 & 0.855 & 0.690 & 0.721 & 0.306 & 0.104 & 0.269 \\
            \cmidrule(lr){1-13}
            ResNet20      & 0.918 & 0.880 & 0.887 & 0.801 & 0.696 & 0.793 & 0.862 & 0.690 & 0.692 & 0.426 & 0.110 & 0.355 \\
            ResNet20 + SE & 0.920 & 0.881 & 0.890 & 0.803 & 0.699 & 0.794 & 0.869 & 0.728 & 0.700 & 0.437 & 0.106 & 0.348 \\
            \cmidrule(lr){1-13}
            ResNet32      & 0.919 & 0.882 & 0.892 & 0.804 & 0.697 & 0.793 & 0.869 & 0.713 & 0.710 & 0.417 & 0.142 & 0.364 \\
            ResNet32 + SE & 0.920 & 0.882 & 0.893 & 0.808 & 0.700 & 0.795 & 0.873 & 0.716 & 0.705 & 0.445 & 0.130 & 0.363 \\
            \cmidrule(lr){1-13}
            ResNet44      & 0.920 & 0.884 & 0.893 & 0.804 & 0.701 & 0.794 & 0.881 & 0.737 & 0.732 & 0.439 & 0.114 & 0.361 \\
            ResNet44 + SE & 0.921 & 0.885 & 0.895 & 0.809 & 0.704 & 0.798 & 0.889 & 0.759 & 0.748 & 0.476 & 0.112 & 0.389 \\
            \cmidrule(lr){1-13}
            ResNet56      & 0.920 & 0.882 & 0.893 & 0.809 & 0.700 & 0.798 & 0.885 & 0.734 & 0.737 & 0.443 & 0.080 & 0.336 \\
            ResNet56 + SE & 0.924 & 0.885 & 0.896 & 0.812 & 0.708 & 0.805 & 0.893 & 0.761 & 0.752 & 0.486 & 0.123 & 0.382 \\
            \bottomrule
        \end{tabular}
    }
\end{table}

\section{Results}
\noindent \textbf{Impact of model architecture.} We evaluated the main architectures used in HE inference and the versions of these models enhanced with SE modules on the CheXpert and NIH datasets. The results are summarized in Table \ref{table1}. For most models, the addition of the SE module resulted in small performance improvements, demonstrating that the SE module worked effectively, even when approximated for HE. In addition, we observed small performance gains in the micro average, which aggregates the results across all classes, indicating an overall improvement in the classification performance. The macro average, which assesses how well the model performs in each class, also showed improved results, suggesting that the model handles all classes effectively and uniformly. Furthermore, the weighted metrics, which consider the sample proportions of each class, showed slight increases in both the F1 and AUC scores. These improvements suggest that the SE module not only improves the overall performance but also ensures balanced performance across different classes.
\begin{table} [t]
    \setlength{\tabcolsep}{4pt} 
    \centering
    \caption{Performance results depending on activation functions. All models in the table were trained and evaluated at $f=8$. $\sigma$ represents sigmoid function. The results represent inferences from unencrypted setup. ReLU/$\sigma$ denote the activations used by typical CNNs, and Poly/Appx $\sigma$ denote the activations approximated by polynomials.}
    \label{table2}
    \centering
    \scalebox{0.8}{
        \begin{tabular}{ccc|cccccc}
            \toprule
            \textbf{CheXpert} & \multicolumn{2}{c}{\textbf{Activation}} & \multicolumn{3}{|c}{\textbf{AUROC} $\uparrow$} & \multicolumn{3}{c}{\textbf{F1} $\uparrow$} \\
            \cmidrule(lr){1-1} \cmidrule(lr){2-3} \cmidrule(lr){4-6} \cmidrule(lr){7-9}
            \textbf{Model} & \textbf{ReLU/$\sigma$} & \textbf{Poly/Appx $\sigma$} & \textbf{Micro} & \textbf{Macro} & \textbf{Weight.} & \textbf{Micro} & \textbf{Macro} & \textbf{Weight.} \\
            \cmidrule(lr){1-9}
            ResNet20      & \checkmark  &  & 0.924 & 0.890 & 0.898 & 0.814 & 0.715 & 0.802 \\
                          &   & \checkmark & 0.918 & 0.880 & 0.887 & 0.801 & 0.696 & 0.793 \\ 
            \midrule
            ResNet20 + SE & \checkmark  &  & 0.926 & 0.894 & 0.901 & 0.816 & 0.720 & 0.808 \\
                          &   & \checkmark & 0.920 & 0.881 & 0.890 & 0.803 & 0.699 & 0.794 \\ 
            \bottomrule
        \end{tabular}
    } 
\end{table}

\begin{table} [t]
    \setlength{\tabcolsep}{4pt} 
    \centering
    \caption{Performance results depending on the inference environment. All models were trained and evaluated at $f=8$. \xmark \ indicates inferences from the unencrypted setup.}
    \label{table3}
    \centering
    \scalebox{0.8}{
        \begin{tabular}{ccc|cccccc}
            \toprule
            \multicolumn{3}{c}{\textbf{CheXpert}} & \multicolumn{3}{|c}{\textbf{AUROC} $\uparrow$} & \multicolumn{3}{c}{\textbf{F1} $\uparrow$} \\
            \cmidrule(lr){1-3} \cmidrule(lr){4-6} \cmidrule(lr){7-9}
            \textbf{Model} & \textbf{ HE } & \textbf{Inference Time} & \textbf{Micro} & \textbf{Macro} & \textbf{Weight.} & \textbf{Micro} & \textbf{Macro} & \textbf{Weight.} \\
            \cmidrule(lr){1-9}
            ResNet20      & \xmark     & 6.23ms     & 0.918 & 0.880 & 0.887 & 0.801 & 0.696 & 0.793 \\ 
                          & \checkmark & 76913.72ms & 0.912 & 0.875 & 0.884 & 0.796 & 0.691 & 0.788 \\ 
            \midrule
            ResNet20 + SE & \xmark     & 8.39ms     & 0.920 & 0.881 & 0.890 & 0.803 & 0.699 & 0.794 \\ 
                          & \checkmark & 92142.15ms & 0.915 & 0.876 & 0.888 & 0.799 & 0.695 & 0.790 \\ 
            \bottomrule
        \end{tabular}
    }
\end{table}

\vspace{3pt}
\noindent \textbf{Impact of activation function.} For HE inference, we approximated the activation functions using polynomials. However, this approximation can lead to a decrease in the performance. To assess the extent of performance changes, we compared and tested two versions of the activation functions for a representative CNN, ResNet20, in an unencrypted setup (Table \ref{table2}). When polynomial activation was used for HE inference, the performance decreased slightly; but, the decrease was not significant. The parameters of the activation function were well-trained, allowing it to function effectively and successfully perform X-ray classification.

\begin{figure} [t]
    \centering
    \includegraphics[width=\textwidth]{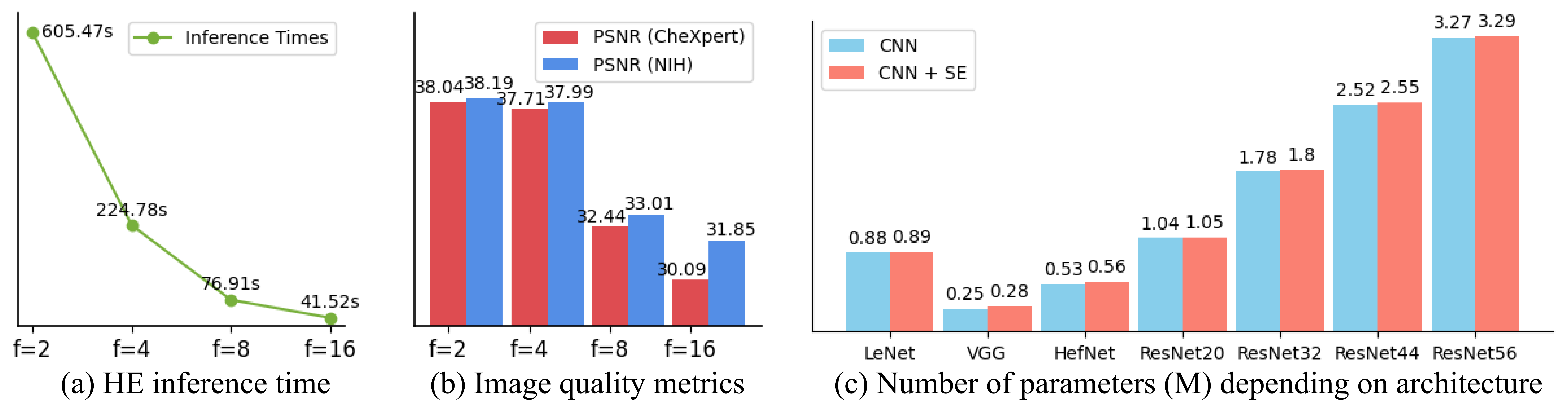}
    \caption{(a) HE inference time for different latent sizes for ResNet20. (b) Reconstructed image quality in PSNR for different latent sizes. (c) Number of parameters for various model architectures based on $f=8$.}
    \label{fig5}
\end{figure}

\vspace{3pt}
\noindent \textbf{Impact of HE inference.} The CKKS scheme supports real number operations but incurs slight errors in the computational results \cite{cheon2017homomorphic}. Consequently, there may be differences compared with unencrypted operations. We performed inference in both the unencrypted and HE setups, and compared the results for a representative CNN, ResNet20 (Table \ref{table3}). The changes in performance owing to the inference environment were minor. However, there was a significant difference in the inference time, which increased exponentially with the number of model parameters and latent size in HE. The inference time as a function of the latent size is shown in Fig. \ref{fig5} (a). When the downsampling factor was set to $f=2$, it took almost 10 minwere required to perform the inference using HE on just one sample. For $f=4$ and $f=8$, the inference time was significantly reduced.

\vspace{3pt}
\noindent \textbf{Impact of latent size.} Reducing the latent size makes model training and validation more efficient, but also increases information loss. The changes in the reconstructed image quality based on latent size are shown in Fig. \ref{fig5} (b). For downsampling factors $f=2$ and $f=4$, the image quality was relatively well preserved. However, starting from $f=8$, the image quality decreased. Similarly, as shown in Fig. \ref{fig3} (a), images with $f=4$ retain some lesion details; however, the details are blurred at $f=8$. To determine whether the loss of information in the latent space leads to a decrease in performance, we validated models with different latent sizes for a representative CNN, ResNet20. The results are summarized in Table \ref{table4}. At $f=8$, the reconstructed images lost some detail, but retained sufficient information to prevent a significant performance decrease. However, at $f=16$, a significant decrease in performance was observed. Therefore, we chose $f=8$ in this study, considering computational cost and inference time.

\begin{table} [t]
    \setlength{\tabcolsep}{5pt} 
    \centering
    \caption{Performance results according to latent size.}
    \label{table4}
    \centering
    \scalebox{0.8}{
        \begin{tabular}{lcc|cccccc}
            \toprule
            \multicolumn{3}{c}{\textbf{CheXpert}} & \multicolumn{3}{|c}{\textbf{AUROC} $\uparrow$} & \multicolumn{3}{c}{\textbf{F1} $\uparrow$} \\
            \cmidrule(lr){1-3} \cmidrule(lr){4-6} \cmidrule(lr){7-9}
            \textbf{Model} & \textbf{Factor} & \textbf{Latent size} & \textbf{Micro} & \textbf{Macro} & \textbf{Weight.} & \textbf{Micro} & \textbf{Macro} & \textbf{Weight.} \\
            \cmidrule(lr){1-9}
            ResNet20      & f=2  & $128\times128\times2$ & 0.921 & 0.888 & 0.896 & 0.810 & 0.719 & 0.804 \\
                          & f=4  & $64 \times64 \times3$ & 0.920 & 0.887 & 0.893 & 0.810 & 0.716 & 0.803 \\
                          & f=8  & $32 \times32 \times4$ & 0.918 & 0.880 & 0.887 & 0.801 & 0.696 & 0.793 \\ 
                          & f=16 & $16 \times16 \times8$ & 0.900 & 0.852 & 0.865 & 0.783 & 0.665 & 0.769 \\
            \midrule
            ResNet20 + SE & f=2  & $128\times128\times2$ & 0.923 & 0.888 & 0.897 & 0.813 & 0.717 & 0.805 \\
                          & f=4  & $64 \times64 \times3$ & 0.921 & 0.888 & 0.896 & 0.810 & 0.719 & 0.804 \\
                          & f=8  & $32 \times32 \times4$ & 0.920 & 0.881 & 0.890 & 0.803 & 0.699 & 0.794 \\ 
                          & f=16 & $16 \times16 \times8$ & 0.911 & 0.868 & 0.881 & 0.791 & 0.681 & 0.780 \\
            \bottomrule
        \end{tabular}
    }
\end{table}

\section{Discussion}
We demonstrate a privacy-preserving framework for HE inference in X-ray image classification. Our study shows that a downsampling factor of eight provides the best balance between computational cost and image quality. However, the computational cost remains a limitation because HE inference is still relatively slow, and the results are slightly different from those of unencrypted inference. However, the errors introduced were minimal and did not significantly affect the overall performance. In addition, image compression must be performed on the client side, which requires the use of a VQGAN encoder on client devices. Although typical CPU resources can handle compression, processing power from the client is required, which can be limited in certain cases. Despite these challenges, our framework has great potential for practical use in medical imaging while ensuring privacy. Recent research has been conducted on medical image analysis using compressed 3D latent \cite{kim2024adaptive,kim2025tumor}. With further improvements in the computational cost, our approach may be effectively applied in real-world settings and 3D medical images.

\begin{credits}
\subsubsection{\ackname} This study was supported by National Research Foundation (RS-2024-00408040), AI Graduate School Support Program (Sungkyunkwan University) (RS-2019-II190421), ICT Creative Consilience program (IITP-2025-RS-2020-II201821), and the Artificial Intelligence Innovation Hub program (RS-2021-II212068).
\end{credits}

{
    \small
    \bibliographystyle{splncs04}
    \bibliography{refs}
}

\end{document}